\documentclass[a4paper]{jpconf}
\usepackage{graphicx}
\begin{document}
\title{Exploring the Nucleon Structure \\ from First Principles of QCD}

\author{W.\ Bietenholz$^{\rm a}$, N.\ Cundy$^{\rm b}$,
M.\ G\"{o}ckeler$^{\rm b}$, R.\ Horsley$^{\rm c}$,
H.\ Perlt$^{\rm d}$, \\ D.\ Pleiter$^{\rm e}$, P.E.L.\ Rakow$^{\rm f}$, 
G.\ Schierholz$^{\rm b,g}$, A.\ Schiller$^{\rm d}$, 
T.\ Streuer$^{\rm b}$ and J.M.\ Zanotti$^{\rm c}$ \ (QCDSF Collaboration)}

\address{$^{\rm a}$ Instituto de Ciencias Nucleares,
Universidad Nacional Aut\'{o}noma de M\'{e}xico, \\
\ \ \ A.P.\ 70-543, C.P.\ 04510 M\'{e}xico, Distrito Federal, Mexico \\
$^{\rm b}$ Institut f\"{u}r Theoretische Physik,
Universit\"{a}t Regensburg, 93040 Regensburg, Germany \\
$^{\rm c}$ School of Physics, University of Edinburgh, 
Edinburgh EH9 3JZ, United Kingdom \\
$^{\rm d}$ Institut f\"{u}r Theoretische Physik,
Universit\"{a}t Leipzig, 04109 Leipzig, Germany\\
$^{\rm e}$ John von Neumann Institut f\"{u}r Computing,
Deutsches Elektronen-Synchrotron DESY, \\
\ \ \ 15738 Zeuthen, Germany \\
$^{\rm f}$ Theoretical Physics Division, Dept.\ of 
Mathematical Sciences, University of Liverpool, \\ 
\ \ \ Liverpool, L69 3BX, United Kingdom \\
$^{\rm g}$ Deutsches Elektronen-Synchrotron DESY, 
22603 Hamburg, Germany}

\ead{wolbi@nucleares.unam.mx}

\begin{abstract}
Quantum Chromodynamics (QCD) is generally assumed to
be the fundamental theory underlying nuclear physics.
In recent years there is progress towards investigating 
the nucleon structure from first principles of QCD.
Although this structure is best revealed in Deep
Inelastic Scattering, a consistent analysis has to be
performed in a fully non-perturbative scheme. The only
known method for this purpose are lattice simulations.
We first sketch the ideas of Monte Carlo simulations
in lattice gauge theory. Then we comment in particular 
on the issues of chiral symmetry and operator mixing.
Finally we present our results for the Bjorken variable
of a single quark, and for the second Nachtmann moment
of the nucleon structure functions.
\end{abstract}

\vspace*{-5mm}

\section{Introduction}

For 37 years we have a strong candidate for a fundamental
theory underlying nuclear physics.
There is a well-founded paradigm that the whole
of nuclear physics can in principle be explained by QCD.
Nevertheless this theory has hardly been represented at this
workshop, because most talks referred to complex phenomena,
which can in practice still not be computed based on QCD.
Effective approaches --- like the shell-model ---
are popular for this purpose. They are often successful for
specific questions, but they all have their limitations.

In this contribution we do address QCD, and we report
in particular on progress in the understanding of the nucleon
structure function from first principles. This is obviously
a non-perturbative task, and the only method to tackle
it systematically are lattice simulations.

The formulation of QCD looks amazingly simple: it deals
with quarks with a local $SU(3)$ symmetry (``colour symmetry''),
which gives rise to strong interaction mediated by 8 gauge bosons
(``gluons''); one for each generator of the gauge group
$SU(3)$. Since the latter is non-Abelian, the gluons
also interact among each other, which makes the dynamics of
the system complicated. In the low energy regime, which dominates
our daily experience, the gauge coupling is strong, and the
interactions are extremely involved. This can be seen for instance
from the fact that the fundamental masses of the lightest (valence) quark 
flavours --- which we assume to be generated by the Higgs mechanism ---
only contribute about 2 $\%$ to the nucleon mass; all the
rest is contained in some dense ``foam'' of gluons and sea-quarks.

So far perturbation theory could be carried out for some quantities 
up to $\alpha_{\rm s}^{4}$, with 
$\alpha_{\rm s} = g_{\rm s}^{2} / 4 \pi$, $g_{\rm s}$ being to
strong gauge coupling. This is applicable
to a number of high energy processes, 
dominated by the exchange of few gluons. 
A suitable method for that purpose is dimensional 
regularisation, which is, however, strictly limited to perturbation
theory.

On the other hand, the lattice regularisation enables a fully
non-perturbative treatment, {\it i.e.} we can really address
finite values of $\alpha_{\rm s}$. Here we refer to the functional
integral formulation of quantum field theory, which is (in principle)
based on an integration over all possible configurations of the fields
involved. The partition function, and all $n$-point functions
that can be derived from it, are then given by an integration over
all these configurations with a phase factor $\exp ({\rm i}S)$, where
$S$ is the action of the configuration. The convergence of such integrals
can be improved drastically by moving to Euclidean space-time,
 {\it i.e.} by invoking an ``imaginary time'' (Wick rotation). This
changes the factor mentioned above to $\exp (-S_{\rm E})$, where 
$S_{\rm E}$ is the Euclidean action. This transition is justified
in view of the expectation values of $n$-point functions if four
conditions --- known as the ``Osterwalder-Schrader axioms'' --- hold.
For many theories, including QCD at zero vacuum angle $\theta$
and zero baryon density, $S_{\rm E}$ is real positive for all
configurations, hence the factor $\exp (-S_{\rm E})$ can be 
interpreted as a probability distribution for the possible field
configurations. 

Lattice studies consider discrete points in a finite volume of
Euclidean space-time. The matter fields --- in this case the 
quark fields $\bar \Psi$, $\Psi$ --- are
located on these lattice sites, while the gauge field variables
$U_{\mu} \in SU(3)$ live on the links connecting nearest neighbour 
sites. Thus the functional integral is completely well-defined.
Most theories of interest are bilinear in the fermion fields. 
Then the action takes the structure
$S_{\rm E}[\bar \Psi , \Psi , U] 
= \bar \Psi M[U] \Psi + S_{\rm gauge}[U]$, where the
indices of $\bar \Psi$, $\Psi$ run over all lattice sites, and
on each site over all internal degrees of freedom. The
fermion matrix $M$ contains the gauge couplings between
neighbouring sites (due to the discrete covariant derivative
in the Dirac operator). $S_{\rm gauge}$ is the pure gauge part of the
action. In accordance with the Spin-Statistics Theorem, the components
of the spinor fields $\bar \Psi$, $\Psi$ are given by Grassmann 
variables, {\it i.e.}\ they anti-commute, 
and the corresponding integration yields
\begin{equation}
Z = \int {\cal D} \bar \Psi {\cal D} \Psi {\cal D}U \,
\exp(-S_{\rm E}[\bar \Psi , \Psi , U])
= \int {\cal D}U \, {\rm det} M[U] \, e^{-S_{\rm gauge}[U]} \ .  
\end{equation}
The functional measure ${\cal D}U$ now represents the integral over
the (compact, gauge invariant) Haar measure for each link variable.
Thus we preserve exact gauge invariance on the regularised level.
%in contrast to other regularisations.
For system sizes of interest, it is still far beyond our
ability to compute this integral explicitly.
Hence the method of choice is to generate a large set of gauge
configurations with the given probability distribution; moreover
the configurations should be well de-correlated, {\it i.e.}\
statistically independent from each other. There are powerful
algorithms for this purpose, see Refs.\ \cite{MLalgo} for reviews.
Once such a set of configurations
is generated, it can be used to measure
observables of interest. This approximate functional
integral yields results for expectation
values with statistical errors (due to the finite sample of
configurations) and systematic errors (finite lattice spacing,
finite volume etc.). These errors can be estimated and reduced
with additional computational effort. The important point,
however, is that the result is fully non-perturbative.
This allows us nowadays for instance to calculate the light 
hadron spectrum from first principles of QCD up to a precision of a 
few percent. The results agree with phenomenology within these
errors. \\

Our daily experience is dominated by nuclear physics at low energy.
Regarding QCD this implies a dominance by the light quark flavours,
in particular the flavours $u$ and $d$. Their masses are very small
compared to $\Lambda_{\rm QCD}$ --- the intrinsic scale of of QCD,
which provides a scale for the hadron masses. Hence these quarks
have an approximate chiral symmetry, which is of prominent 
importance in low energy QCD. To briefly review this property,
we first note that fermion fields can be decomposed
into ``left- and right-handed'' components by means of the chiral
projectors,
\begin{equation}
\Psi_{\rm L,R} = \frac{1}{2} ( 1 \pm \gamma_{5}) \, \Psi \ , \quad
\bar \Psi_{\rm L,R} = \bar \Psi \, \frac{1}{2} ( 1 \mp \gamma_{5}) \ .
\end{equation}
In the chiral limit (zero quark masses) these two components
decouple. It is allowed to add quark masses to the Lagrangian,
since QCD is a vector theory --- the left- and right-handed components
couple in the same way to the gauge field --- (unlike
the electroweak sector of the Standard Model).
From a perturbative perspective, the approximate chiral symmetry
still protects the masses of light quarks
from large renormalisation. Based on this
property, it is occasionally claimed that there is no hierarchy problem
for fermion masses. However, it is the non-perturbative level that
ultimately matters for physics, and there it
is highly non-trivial to implement (approximate) chiral symmetry
in a regularised system. A quite obvious ansatz is the Wilson lattice 
fermion, which starts from the na\"{\i}ve discretisation of the
derivative in the Dirac operator, and subtracts a discrete Laplacian
to send the fermion doublers to the cutoff scale. However, this extra
term breaks chiral symmetry explicitly. As a
consequence, under gauge interaction the fermion masses are 
renormalised such that they naturally end up
at the cutoff scale as well. The (observed) property 
$m_{u,d} \ll \Lambda_{\rm QCD}$ can then only be attained by a
tedious fine-tuning of negative bare quark masses, {\it i.e.}\
the hierarchy problem is back.

Only at the very end of the last century was there a breakthrough
in the formulation of chiral lattice fermions, which overcomes
this problem, at least in vector theories.
A key observation was that a lattice
modified version of chiral symmetry of the form \cite{ML98}
\begin{equation}
\bar \Psi \to \bar \Psi \Big( 1 + \varepsilon \Big[ 1 -
\frac{a}{2} D \Big] \gamma_{5} \Big) \ , \quad
 \Psi \to \Big( 1 + \varepsilon \gamma_{5} \Big[ 1 -
\frac{a}{2} D \Big] \Big) \Psi \qquad 
({\rm to} \ O( \varepsilon )) \ ,
\end{equation}
is sufficient to rule out additive mass renormalisation. 
Here $a$ is the lattice spacing, and in the continuum limit this
turns into the standard chiral transformation (infinitesimal form).
Even on the lattice the fermionic Lagrangian $\bar \Psi D \Psi$
is invariant, {\em if} the lattice Dirac operator $D$ fulfils the
Ginsparg-Wilson relation (GWR)
\begin{equation} \label{GWR}
\{ D , \gamma_{5} \} \equiv D \gamma_{5} + \gamma_{5} D
= a D \gamma_{5} D \ .
\end{equation}
On the other hand, the functional measure 
${\cal D} \bar \Psi {\cal D} \Psi$ is not invariant, and this is
just what it takes to reproduce the axial anomaly correctly 
\cite{ML98}. 

Solutions to the GWR can be constructed based on block spinor
renormalisation group transformations (``perfect lattice
fermion'') \cite{GiWi,Has98}, or by
pulling apart the chiral modes in an extra ``dimension''
(``Domain Wall Fermion'') \cite{Kaplan}. 
By integrating out that ``dimension'', one obtains ---
for $\gamma_{5}$-Hermitian operators, $D^{\dagger} = \gamma_{5} D
\gamma_{5}$ --- the ``overlap Dirac operator'' $D_{\rm ov}$ \cite{Neu},
\begin{equation}
a D_{\rm ov} = 1 + A_{\rm ov} \ , \quad 
A_{\rm ov} = A / \sqrt{ A^{\dagger} A} \ , \quad A = a D -1 \ . 
\end{equation}
It is easy to verify that it satisfies the GWR, which can now be
written as 
$D_{\rm ov} + D^{\dagger}_{\rm ov} = a D^{\dagger}_{\rm ov} D_{\rm ov}$.
For the kernel $D$ we can insert for instance the Wilson operator
 \cite{Neu},
but properties like scaling, locality and approximate rotation
symmetry can be improved by using a better kernel \cite{WB99}.

%Due to the exact, lattice modified chiral symmetry, this operator
%also has exact zero modes with a definite chirality, and therefore
%a well-defined fermion index. This index can be used to define a
%topological charge of the lattice gauge configurations 

In addition to the exact, lattice modified chiral symmetry, the GWR
also prevents uncontrolled operator mixing on the regularised level
\cite{Has98}.
This powerful property is very helpful, for instance for the numerical 
study of a fully non-perturbative Operator Product Expansion (OPE). \\

We now summarise our lattice OPE project \cite{OPE1,OPE2,OPE3} as an example 
for the application of chiral quarks in lattice QCD. This takes us beyond 
%the evaluation of 
the hadron spectrum; we investigate the internal structure
of the nucleon (see Ref.\ \cite{Haeg} for a recent review). 
Phenomenologically this structure is revealed
in Deep Inelastic Scattering %(``Deep Inelastic Scattering''), 
of a lepton on a nucleon target. This scattering is dominated by a 
one-photon exchange involving a single quark, as sketched below.
\begin{figure}[h!]
\begin{center}
\vspace*{-2mm}
\includegraphics[angle=0,width=.4\linewidth]{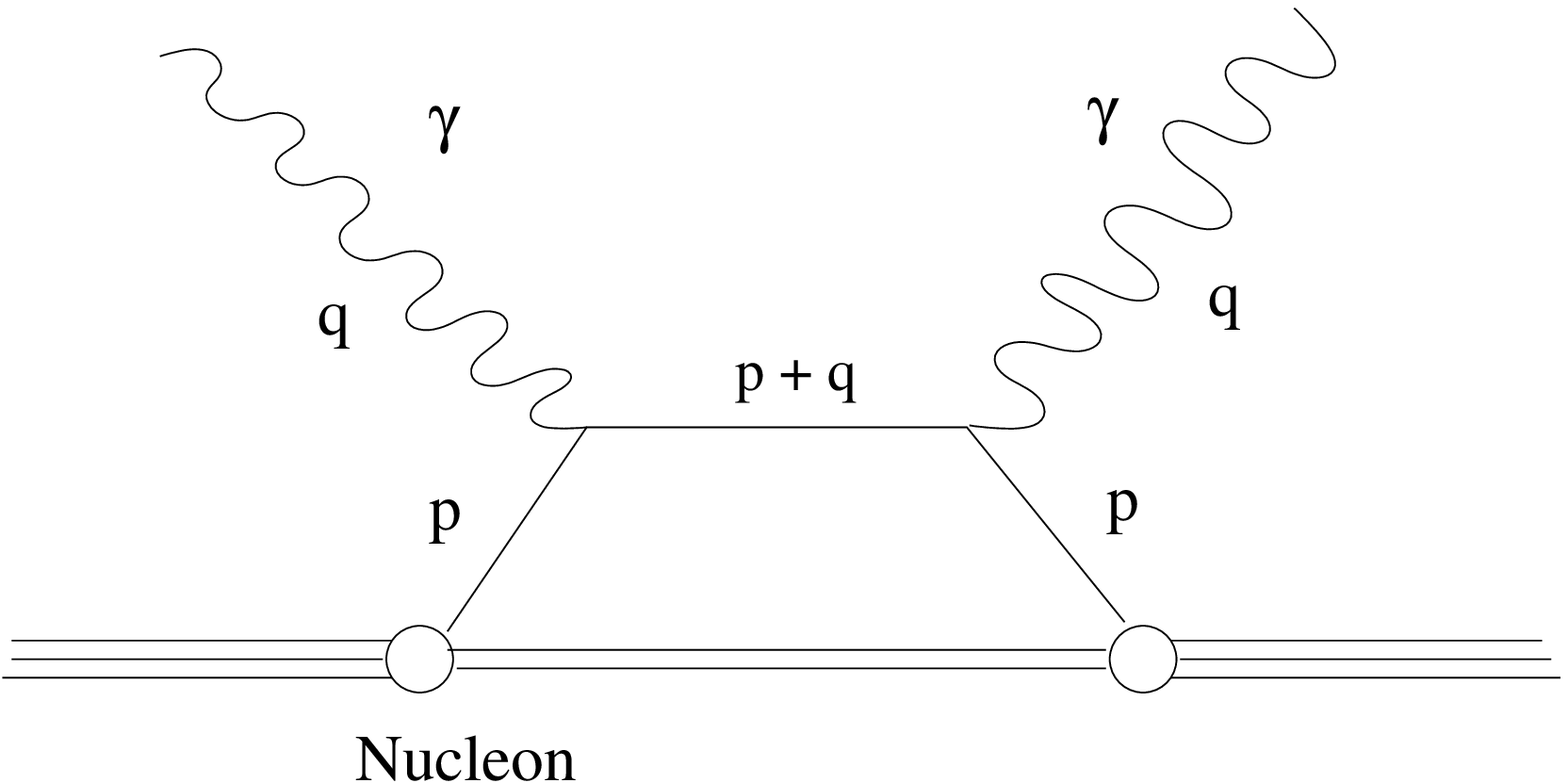} \hspace*{5mm}
\includegraphics[angle=0,width=.4\linewidth]{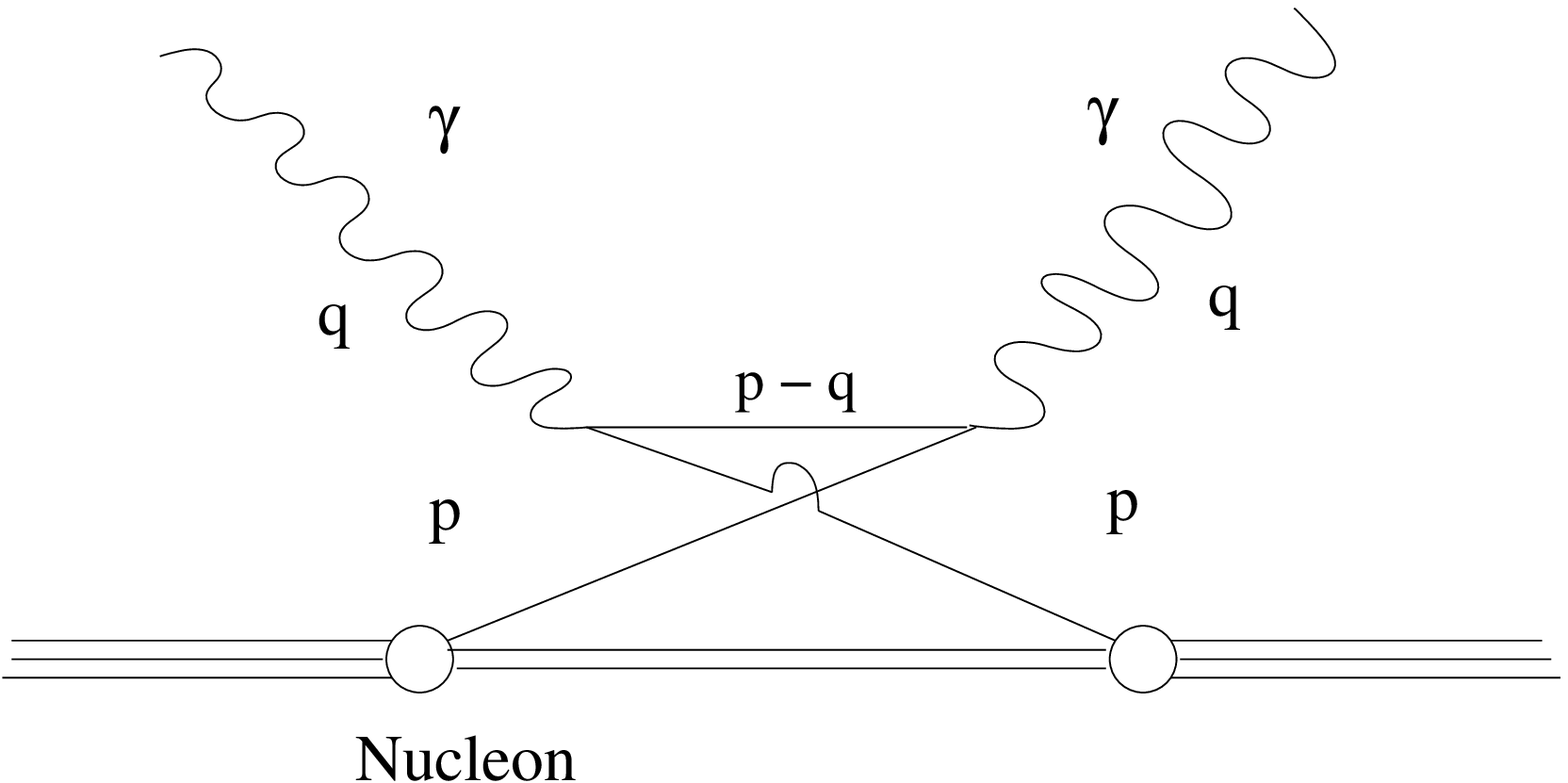}
\end{center}
\vspace*{-4mm}
\end{figure}

Although this is a high energy process, there are problems with
a perturbative treatment due to ambiguous UV and IR divergences.
%, which give rise to renormalon ambiguities. 
A fully satisfactory study based on QCD
has to be consistently non-perturbative, {\it i.e.}\ it is a
challenge for the lattice. The involvement of a single quark
implies that chirality is conserved, hence the use of Ginsparg-Wilson
quarks is appropriate. In our simulations we applied 2 flavours of
degenerate overlap valence quarks with masses of 29~MeV and 73~MeV, 
which correspond to pion masses of 280 MeV and 440 MeV, respectively.
The goal is the evaluation of the product of electromagnetic
currents $J_{\mu}$ --- of the photon exchange, with large
transfer momentum $q$ --- between quark states of low momentum $p$.
This product is decomposed by the OPE as follows,
\begin{equation} \label{OPEsum}
W_{\mu \nu} \simeq \langle \Psi (p) | J_{\mu}(q) J_{\nu}(-q) |
\Psi (p) \rangle = \sum_{m,i,n} C^{(m)}_{\mu \nu ,i,\mu_{1} \dots
\mu_{n} }(q) \langle \Psi (p) | {\cal O}^{(m)}_{i,\mu_{1} \dots
\mu_{n} } | \Psi (p) \rangle \ .
\end{equation}
$C^{(m)}$ are Wilson coefficients, which are independent of the
target and therefore of $p$, and ${\cal O}^{(m)}$ are {\em local} 
operators,  which are relevant to describe the nucleon structure. 
They involve the quark momentum components $p_{\mu_{j}}$. We further
use the Clifford index $i=1 \dots 16$, and the index $m$ to distinguish
operators with the same symmetries.

\begin{figure}[h!]
\begin{center}
\includegraphics[angle=270,width=.45\linewidth]{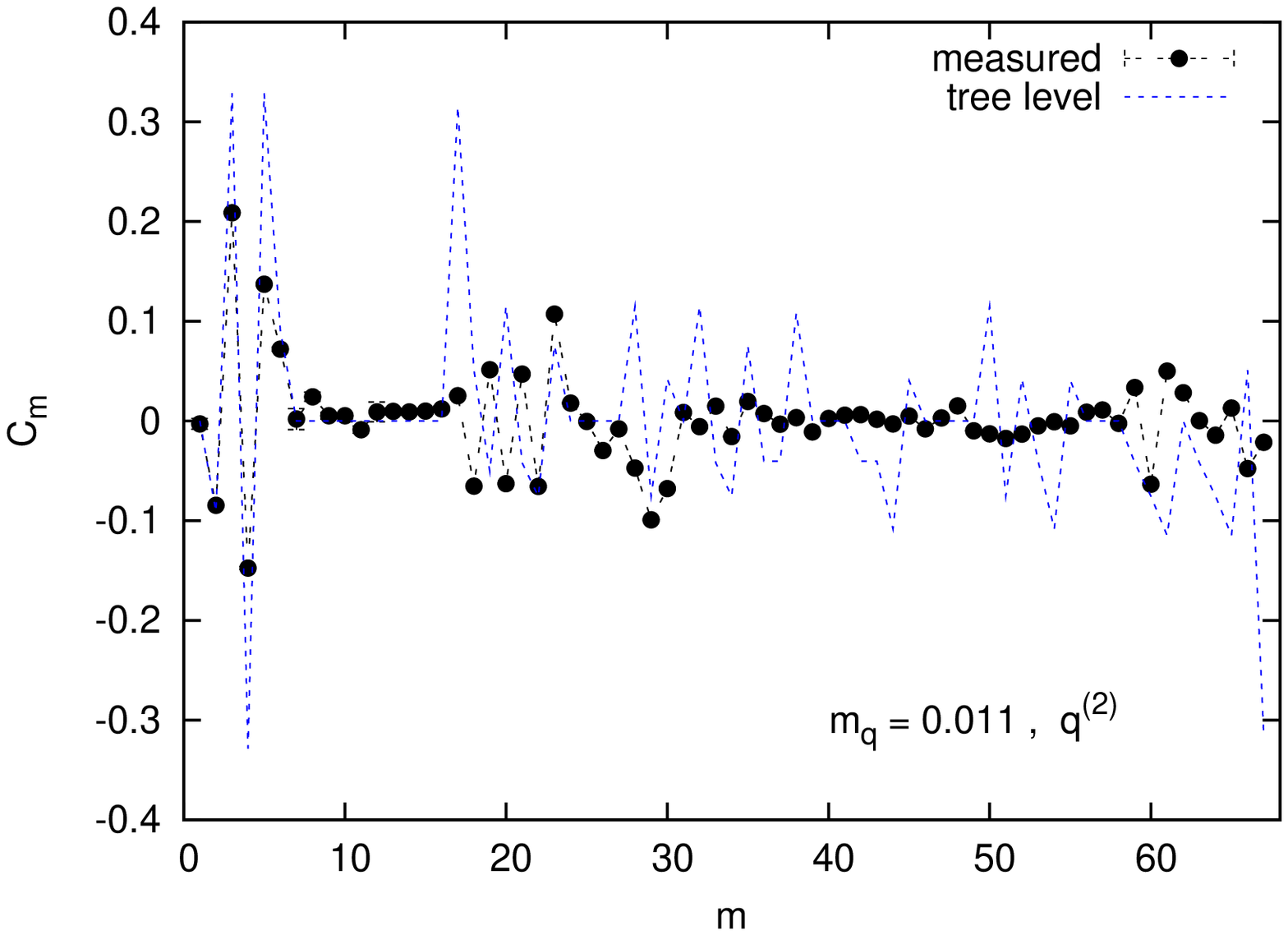} 
%\hspace*{5mm}
\includegraphics[angle=270,width=.45\linewidth]{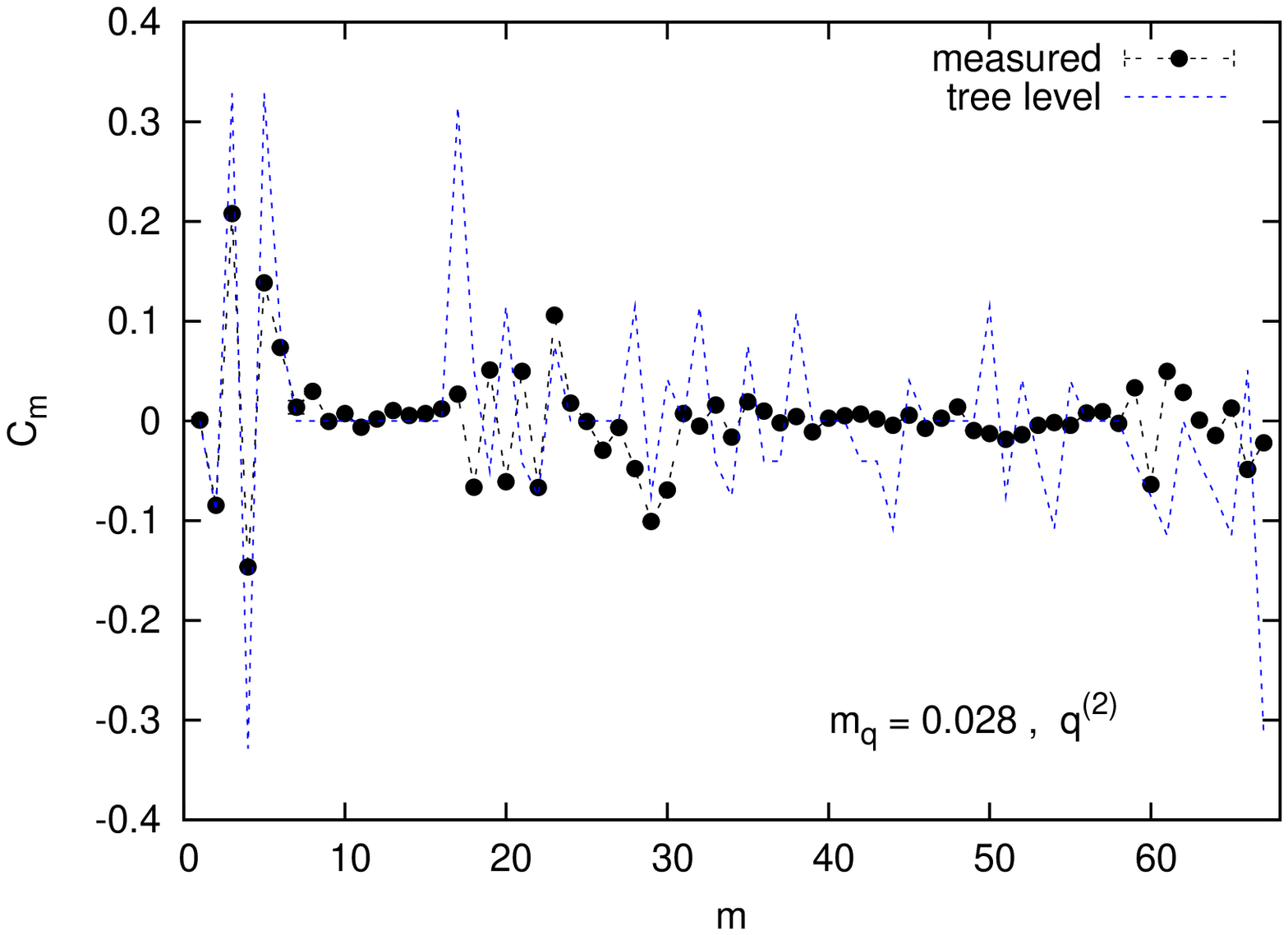}
\end{center}
\vspace*{-3mm}
\label{WcoefFig}
\caption{\small The Wilson coefficients for quark masses $m_{q}=0.011$ and
 $m_{q}=0.028$ in lattice units, corresponding to 29 MeV and 73 MeV.
%respectively. 
In both cases the photon momentum amounts to 3.3 GeV.
The results for the 67 independent coefficients are compared to their
values on tree level. We see a similar pattern, with reduced absolute
values under interaction. The coefficients which vanish in the chiral 
limit, $C_{1}$ and $C_{7} \dots C_{16}$, are still
small at finite quark mass, which confirms the validity of our method.}
\vspace*{-4mm}
\end{figure}

A truncation of the OPE on the right-hand-side of eq.\ (\ref{OPEsum})
%, and small lattice artifacts, 
requires the scale separation 
\begin{equation}
p^{2} \ll q^{2} %\ll (\pi / a)^{2} 
\ .
\end{equation}
We probed three photon momenta, $|q^{(1)}| = 2.2 ~ {\rm GeV}$, 
 $|q^{(2)}| = 3.3 ~ {\rm GeV}$, $|q^{(3)}| = 4.4 ~ {\rm GeV}$.
Based on the above scale separation, we truncated the quark
bilinears at $O (|p|^{3})$. Thus we still include 1360 operators, but
for photon momenta $q \propto (1,1,1,1)$ they occur in only 67
equivalence classes. The strategy is now to measure the elements
of $W_{\mu \nu}$ and $\langle {\cal O}^{(m)} \rangle$ for a
multitude of low quark momenta $p$, and solve eq.\ (\ref{OPEsum}) for
the Wilson coefficients. They are over-determined in this
way, but by means of a Singular Value Decomposition we arrived
at reliable values \cite{OPE1}. 
In Fig.\ \ref{WcoefFig} we compare the results obtained numerically at 
photon momentum $q^{(2)}$ to the Wilson coefficients at tree level
in the chiral limit. We see that they follow the same pattern, though
gauge interaction reduces the absolute values significantly. 
In particular the coefficients
$C_{1}$ and $C_{7} \dots C_{16}$ vanish in the chiral limit, because they
are attached to operators with an even power of $\gamma$-matrices.
We obtain small absolute values for them also in the interacting
case at finite quark mass, which confirms that chiral symmetry on the
lattice works, and it suppresses operator mixing. \\

As a first application, we evaluate the Bjorken variable $x$ for one
unpolarised quark. The expectation value 
$\langle x \rangle$ measures the nucleon momentum fraction
carried by the quark involved in the Deep Inelastic Scattering.
To be explicit, we consider the tensor \cite{OPE2}
\begin{equation}
T_{\mu \nu} = \frac{1}{4}{\rm Tr} [ S^{-1} (p) W_{\mu \nu} (p,q)]
= \delta_{\mu \nu} W_{1} + p_{\mu} p_{\nu} W_{2} +
(p_{\mu} q_{\nu} + p_{\nu} q_{\mu}) W_{4} + q_{\mu} q_{\nu} W_{5} \ ,
\end{equation}
where $S$ is the quark propagator. The Dirac-index structure is not
visible here due to the trace (terms with $W_{3}$ and $W_{6}$ would
occur for polarised targets). 

\begin{figure}[h!]
\begin{center}
\includegraphics[angle=270,width=.43\linewidth]{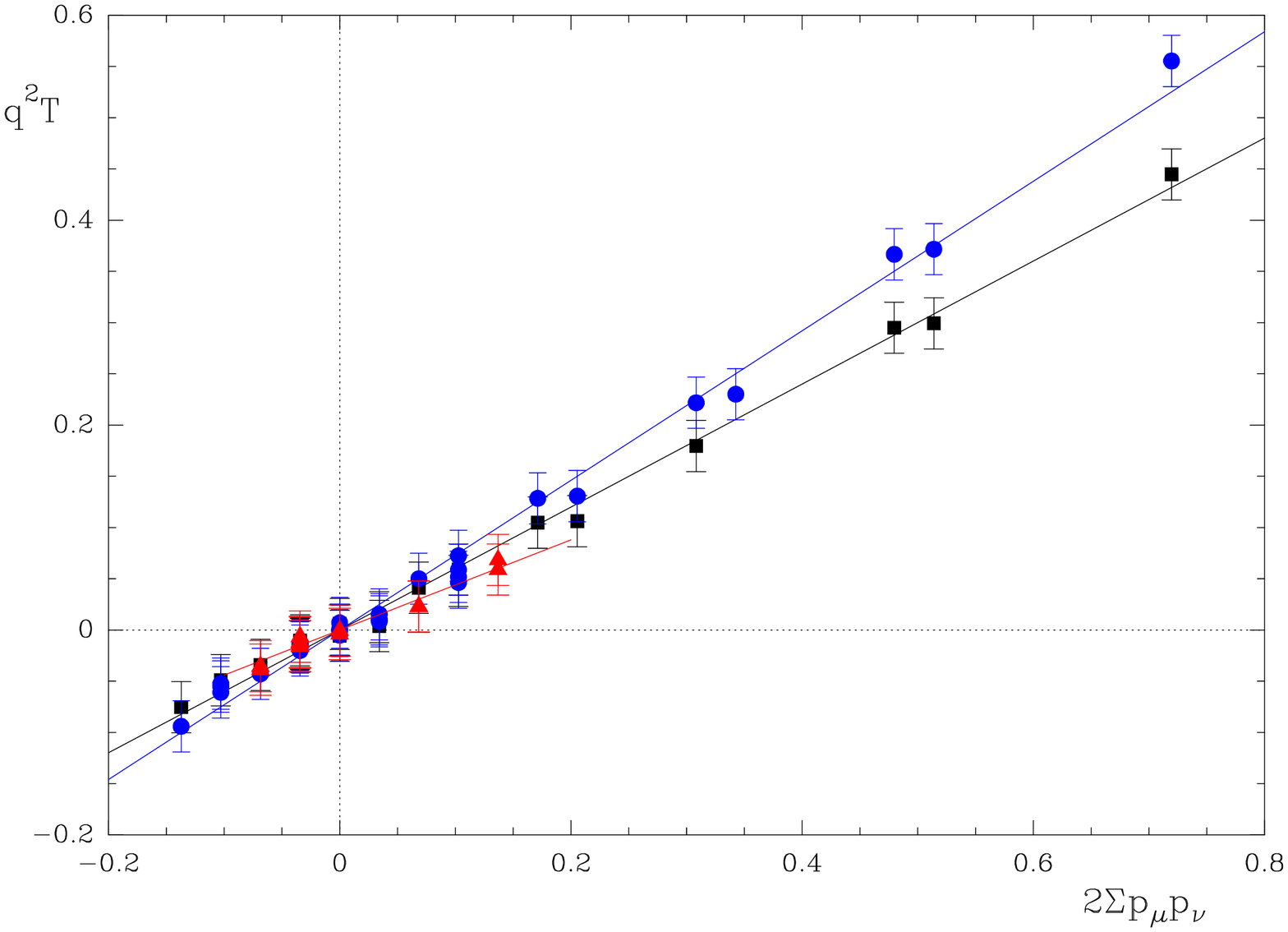} 
%\hspace*{5mm}
\includegraphics[angle=270,width=.49\linewidth]{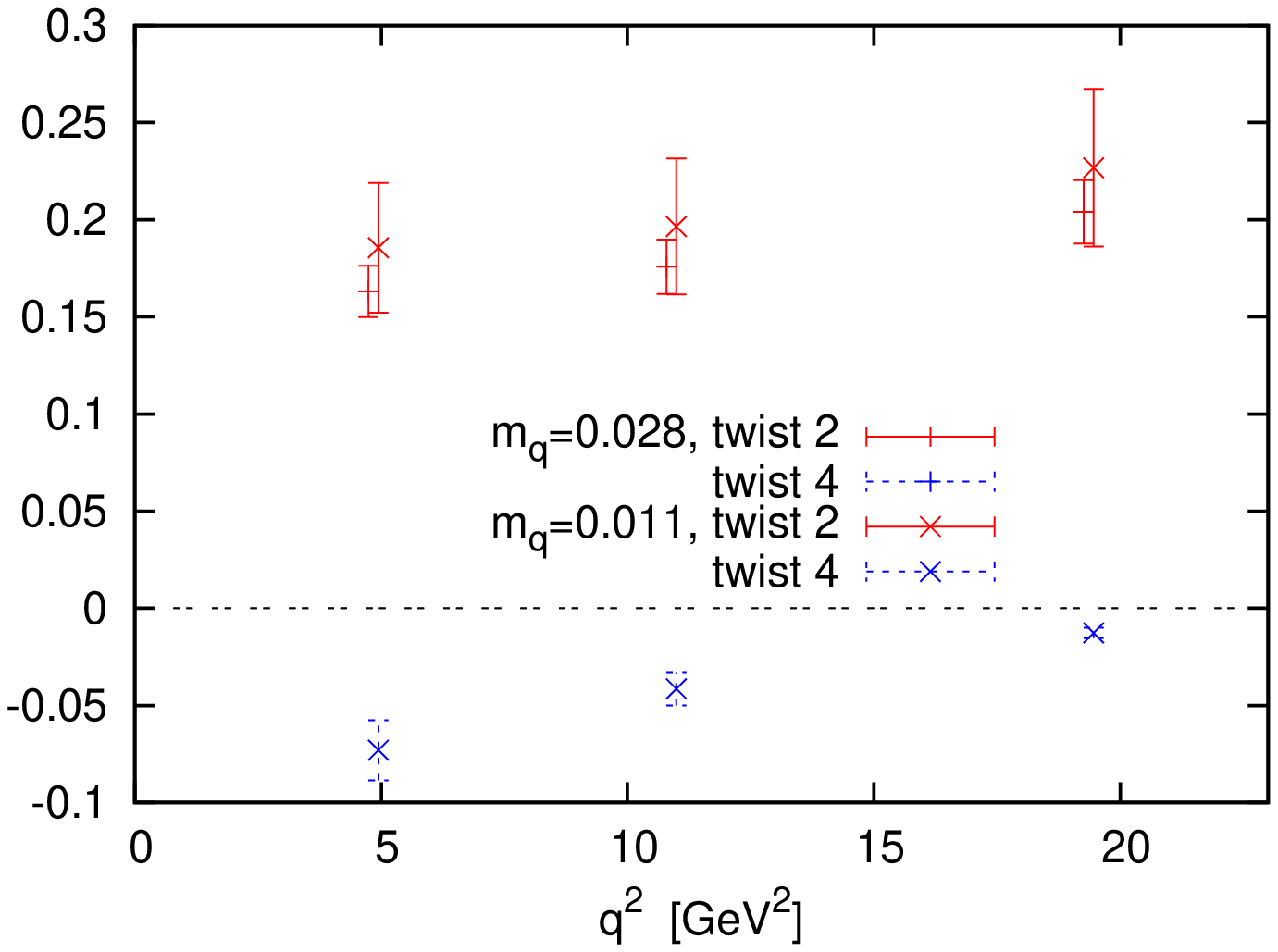}
\end{center}
\vspace*{-5mm}
\caption{\small {\it On the left:} The term $q^{2} T_{\mu \mu}$ for spin 2,
versus $2 \sum_{\mu > \nu} p_{\mu} p_{\nu}$ (numerator and denominator 
in eq.\ (\ref{Bjox})). The squares, circles and triangles refer to
photon momentum $q^{(1)}$, $q^{(2)}$, $q^{(3)}$, respectively.
In all three cases the data points are compatible with a linear
behaviour. Thus the slope allows for the determination of
the Bjorken variables $\langle x \rangle$ given eq.\ (\ref{xtab}).
{\it On the right:} The nucleon moment ${\cal M}_{2}$ against the photon
momentum squared. We distinguish the contributions of twist 2
and twist 4. In all cases the result for our two quark masses
agree within statistical errors. For large
$q^{2}$ the twist 4 contribution seems to vanish, whereas the
twist 2 contribution stabilises at a finite value
in the range $\approx 0.5 \dots 0.8$.}
\label{xM2} 
\vspace*{-3mm}
\end{figure}

In particular $T_{\mu \mu}$ is directly related to the Bjorken variable
in the high energy Bjorken limit. In this limit, the Minkowski space 
momenta obey $(p \cdot q)^{2} \gg p^{2} q^{2}$. 
Among the 22 operators in the expansion of $T_{\mu \mu}$ we focus on 
those with spin 2, where the Bjorken limit leads to \cite{OPE2}
\begin{equation} \label{Bjox}
\langle x \rangle = \frac{q^{2} \ [T_{\mu \mu}]_{\rm spin~2} }
{2 \, \sum_{\mu > \nu} p_{\mu} p_{\nu} } \ .
\end{equation}
Fig.\ \ref{xM2} (on the left) 
shows the numerator against the denominator of the
right-hand-side in eq.\ (\ref{Bjox}). A broad range of
points measured at $m_{q} = 73 ~ {\rm MeV}$ is consistent with
a linear behaviour, which allows us to read off $\langle x \rangle$
from the slope (the errors are statistical only),
\begin{equation}  \label{xtab}
\langle x \rangle |_{q^{(1)}} = 0.44(9) \ , \quad
\langle x \rangle |_{q^{(2)}} = 0.73(5) \ , \quad
\langle x \rangle |_{q^{(3)}} = 0.60(5) \ .
\end{equation}
%where the errors are statistical only.

Finally we proceed to the nucleon structure function \cite{OPE3}.
On a lattice of spacing $a$, a generic moment can be expanded as
\begin{equation}
{\cal M} (q^{2}) = c_{2}(q^{2}a^{2}) A_{2} (a^{2} ) +
\frac{c_{4}(q^{2}a^{2})}{q^{2}} A_{4} (a^{2} ) + \dots
\end{equation}
where $A_{n}$ are reduced nucleon matrix elements (the Lorentz
structure is factored out), $c_{n}$ are the corresponding Wilson
coefficients, and $n$ is the twist. Both factors are bare
quantities. If one tries to replace $a$ by some scale $\mu$
of a soft renormalisation scheme, one runs into ambiguities,
which do not occur on the lattice with its hard momentum 
cutoff $\pi / a$.

In our study, we evaluated the (spin averaged) nucleon matrix
elements for all 67 types of operators, and extracted moments
of the structure function by Nachtmann integration of the hadronic
tensor \cite{Nacht}.
This allows us in particular to project out operators of spin 2.
For instance, the traceless form of $W_{\mu \nu}$ leads to
\begin{equation}
{\cal M}(q^{2}) = \frac{3 q^{2}}{(4 \pi )^{2}} \int d \Omega_{q}
\ n_{\mu} \Big( W_{\mu \nu} - \frac{1}{4} 
\delta_{\mu \nu} W_{\lambda \lambda} \Big) n_{\nu}
\ \to \ \int_{0}^{1} dx \ \Big( F_{2} (x, q^{2}) + \frac{1}{6}
F_{L} (x, q^{2}) \Big) \ .  \label{F2FL}
\end{equation}
Here $n$ is a constant 4-vector with $n^{2}=1$, and the last
expression in eq.\ (\ref{F2FL}) refers again to the Bjorken limit. 
By means of a further projection \cite{OPE3} we can separate the 
structure functions $F_{2}$ and $F_{L}$. Based on the numerical
measurement of the matrix elements, we can carry out the first integral 
in eq.\ (\ref{F2FL}). Thus we arrive at
explicit and non-perturbative results for ${\cal M}$.

Our results for the second moment ${\cal M}_{2}$ 
are shown in Fig.\ \ref{xM2} (on the right). 
They agree within the errors
for both quark masses, at all three photon momenta, which suggests
that they may be valid up to the physical pion mass. As we increase
the photon momentum ({\it i.e.}\ the transfer momentum in the
hard scattering), the twist 4 contribution tends to vanish,
while the twist 2 term stabilises at a moderate value. 
As usual the numerical errors increase for lighter quarks.
%(they are mostly due to the matrix elements). 

These results demonstrate that the explicit determination of
the nucleon structure functions --- as observed in Deep Inelastic
Scattering --- directly from QCD {\em is} feasible, in the framework 
of a non-perturbative regularisation free of ambiguities.
Moreover we obtained explicit results for the Bjorken variable
of the quark involved in this scattering. Our results
provide insight into the nucleon structure. For
a detailed discussion we refer to Ref.\ \cite{OPE3}.
\vspace*{2mm}

{\small {\bf Acknowledgments :} W.B.\ thanks the organisers of the
interesting and pleasant workshop 
in Cocoyoc. This work was supported by the {\it Deutsche
Forschungsgemeinschaft} (DFG) through project FOR 465 and 
{\it Sonderforschungsbereich} SFB/TR55. The computations were performed
on clusters of the {\it Norddeutscher Verbund f\"{u}r Hoch- und
H\"{o}chstleistungsrechnen} (HLRN).

\end{document}